\documentclass[epj]{webofc}
\usepackage[utf8]{inputenc}
\usepackage[varg]{txfonts}   
\usepackage{booktabs}
\usepackage{xcolor}
\definecolor{darkred}{rgb}{0.4,0.0,0.0}
\definecolor{darkgreen}{rgb}{0.0,0.4,0.0}
\definecolor{darkblue}{rgb}{0.0,0.0,0.4}
\usepackage[bookmarks,linktocpage,colorlinks,
    linkcolor = darkred,
    urlcolor  = darkblue,
    citecolor = darkgreen]{hyperref}
\usepackage{subfigure}
\usepackage{cancel}
\usepackage{tikz}
\wocname{EPJ Web of Conferences}
\woctitle{Lattice2017}
\begin{document}
%
\selectlanguage{english}
\rightline{MS-TP-17-16}
\title{%
Non-perturbative determination of improvement $b$-coefficients \\ in $N_f=3$\thanks{Talk given at the 35th International Symposium on Lattice Field Theory, 18-24 June 2017, Granada, Spain.}
}
\author{%
\firstname{Giulia~Maria} \lastname{de~Divitiis}\inst{1,2}\fnsep\thanks{Speaker, \email{giulia.dedivitiis@roma2.infn.it}} \and
\firstname{Maurizio} \lastname{Firrotta}\inst{1,2} \and
\firstname{Jochen} \lastname{Heitger}\inst{3} \and
\firstname{Carl~Christian} \lastname{K\"oster}\inst{3} \and
\firstname{Anastassios} \lastname{Vladikas}\inst{2} 
}
\institute{%
Dipartimento di Fisica, Universit\`a di Roma Tor Vergata, Via della Ricerca Scientifica 1, 00133 Rome, Italy
\and
INFN, Sezione di Tor Vergata, c/o Dipartimento di Fisica, Universit\`a di Roma Tor Vergata, Via della Ricerca Scientifica 1, 00133 Rome, Italy
\and
Institut f\"ur Theoretische Physik, Universit\"at M\"unster, Wilhelm-Klemm-Str. 9, 48149 M\"unster, Germany
 }
\abstract{%
We present our preliminary results of the non-perturbative determination of the valence mass dependent  
coefficients $b_\mathrm{A}-b_\mathrm{P}$ and $b_\mathrm{m}$ as well as the ratio $Z_\mathrm{P} Z_\mathrm{m}/ Z_\mathrm{A}$ entering 
the flavour non-singlet  PCAC relation in lattice QCD with $N_f=3$ dynamical flavours.  We apply the  method proposed 
in the past for quenched approximation and $N_f=2$ cases, employing a set of finite-volume ALPHA configurations 
with Schr\"odinger functional boundary conditions, generated with $O(a)$ improved Wilson fermions and the tree-level Symanzik-improved
gauge action for a range of couplings relevant for simulations at lattice spacings of about $0.09 \,$fm and below.
}
\maketitle
%
\section{Introduction}
\label{sec:intro}
%
Discretisation effects of lattice quantities computed with Wilson fermions are linear in the lattice spacing $a$,
and may be a source of significant systematic errors, resulting in poor control of the continuum extrapolations of physical observables.
In the Symanzik improvement programme
these $O(a)$ effects can be removed by adding irrelevant operators both to the lattice action
and to the local operators inserted in bare correlation functions. These so-called Symanzik counterterms have coefficients which must be tuned non-perturbatively, in order to remove all $O(a)$ contributions from physical quantities.
The improvement coefficients which multiply mass dependent Symanzik counterterms are referred in the literature as $b$-coefficients. 
We will present preliminary results for the $b$-coefficients related to the renormalised quark masses in QCD with three dynamical
sea quarks. For analogous results on the renormalisation and improvement of the vector current see Ref.~\cite{hjvw}. 
%
\section{Improvement condition}
\label{sec:improvement}
%
The improvement coefficients are short distance quantities. They can be determined by imposing suitable conditions in  small physical volumes.
We adopt the Schr\"odinger functional setup, with $L^3\times T$ lattices having periodic (Dirichlet) boundary conditions
in space (time). The renormalisation scale is $\mu=1/L$. 
As we will exploit the freedom to keep sea- and valence-quark masses distinct,
our setup is non-unitary.
Sea quark masses are tuned to the chiral limit, in line with the usual ALPHA choice of a mass-independent renormalisation scheme.
As the bare coupling $g_0$ is varied, all other
bare parameters (such as the valence quark masses) are tuned so as to stay on a line of constant physics. This ensures that the $b$-coefficients are smooth functions of $g_0$.

 
The non-pertubative definition of the $b$-coefficients is not unique and depends upon the chosen improvement condition. 
The one we use
is the standard non-singlet PCAC relation among renormalised quantities~\cite{Luscher:1996ug}:
\begin{align}   
& {\tilde\partial_{\mu}\left\langle  {A_{R}}_\mu^{ij}(x)   \;{\mathcal{O}}^{ji}  \right\rangle = (m_{\mathrm{R},i}+m_{\mathrm{R},j})  \;\left\langle  {P_{R}}^{ij} (x)   \;{\mathcal{O}}^{ji}  \right\rangle + {O(a^2)}} \, ,
\label{eq:ren-wi}
\end{align}
where ${A_{R}}_\mu^{ij}, \, {P_{R}}_\mu^{ij},$ $m_{\mathrm{R},i}, \, m_{\mathrm{R},j}$ denote the renormalised  axial current, pseudoscalar density and masses with  flavour indices  $i, \, j$. In the following, quantities with the same flavour index, such as $A_\mu^{11}, m_{22}$ etc., are intended as defined for two distinct but degenerate valence flavours, so as to avoid Wick contractions that give rise
to diagrams with disconnected quark lines.
Improvement enforces this Ward identity, which holds in the continuum, to have no corrections linear in the lattice spacing,
extending its validity up to $O(a^2)$ violations. 
Starting from the bare quantities
\begin{align}   
&A_\mu^{ij} \equiv \bar \psi_i \gamma_\mu\gamma_5 \psi_j\,,\quad
 P^{ij} \equiv \bar \psi_i \gamma_5 \psi_j  \, ,\nonumber\\  
 &m_{q,ij} \equiv \tfrac{1}{2} (m_{q,i}+m_{q,j})\, ,\quad m_{q,i} \equiv m_{0,i} - m_{crit}=\tfrac{1}{2a} (\tfrac{1}{\kappa_i}-\tfrac{1}{\kappa_{crit}})  \, ,
\label{eq:bare-quantities}
\end{align}
we can write the renormalised masses and operators, in standard notation, as~\cite{Bhattacharya:2005rb}:
\begin{align}   
{A_{R}}_\mu^{ij}&=
 Z_\mathrm{A} \,(1+ {b_\mathrm{A} \,}a m_{q,ij}  
+\cancel{{\bar{b}_\mathrm{A}}\, a\, \mathrm{tr} \,\hat m^{({\rm sea})}})\,
 \{A_\mu^{ij}+{c_\mathrm{A}} a\, \tilde{\partial}_\mu
       P^{ij}\}  \, , \nonumber\\
{P_{R}}_\mu^{ij}&=
Z_\mathrm{P} \,(1+ {b_\mathrm{P}} \,am_{q,ij} +\cancel{{\bar{b}_\mathrm{P}}\, a\,\mathrm{tr} \,\hat m^{({\rm sea})}}) \, P^{ij}  \, , \label{eq:ren-quantities}\\
  m_{\mathrm{R},i} &= Z_\mathrm{m} \Big \{ m_{q,i} (1+{ b_\mathrm{m}}\, a m_{q,i} + \cancel{{\bar{b}_\mathrm{m}}\, a\, \mathrm{tr} \,\hat m^{({\rm sea})}})
+ \cancel{{x}\, \mathrm{tr} \,\hat m^{({\rm sea})}} + \cancel{{y}\,a \,\mathrm{tr} \, \hat m^{2({\rm sea})}} + \cancel{{z}\, a\, (\mathrm{tr} \,\hat m^{({\rm sea})}})^2\Big \} \,. \nonumber\\ 
& {\scriptstyle \hspace{6.3 cm} x\equiv (1-r_\mathrm{m})/N_f 
\quad y\equiv (r_\mathrm{m} d_\mathrm{m}- b_\mathrm{m})/N_f
\quad z\equiv (r_\mathrm{m}\bar d_\mathrm{m}-\bar b_\mathrm{m})/N_f}\nonumber
\end{align}
In small print we give the  expressions  for $x,y$ and $z$ in terms of the parameters $r_\mathrm{m}, b_\mathrm{m}, \bar b_\mathrm{m}, d_\mathrm{m}, \bar d_\mathrm{m} $ defined in Ref.~\cite{Bhattacharya:2005rb}.
It is important to keep in mind that the coefficients $b_\mathrm{A}, b_\mathrm{P}$ and $b_\mathrm{m}$, 
multiplying valence quark masses, arise from the mass dependence of the valence quark propagators 
and contain also mass-independent contributions from the fermion loops. On the other hand
$\bar b_\mathrm{A},\bar b_\mathrm{P},x,y,z$  arise from the mass dependence of quark fermion loops. By keeping valence and sea quark masses distinct and tuning the bare (subtracted) sea-quark mass-matrix $\hat m^{({\rm sea})}$ to the chiral limit, the above expressions simplify as indicated.
%
\section{Non-perturbative definitions of $b_\mathrm{A}-b_\mathrm{P}$, $b_\mathrm{m}$, and $Z$}
\label{sec:NPdefinitions}
%
We compute Schr\"odinger functional correlation functions
\begin{align}   
& f^{ij}_\mathrm{A}(x_0)       \equiv -{a^3}\sum_{\mathbf{x}}\big\langle A^{ij}_0(x)\, {\mathcal{O}^{ji}}  \big\rangle  \, ,\nonumber\\
 &f^{ij}_\mathrm{P}(x_0)       \equiv -{a^3}\sum_{\mathbf{x}}\big\langle P^{ij}(x)\, {\mathcal{O}^{ji}} \big\rangle  \, ,\qquad  \qquad  \mathcal{O}^{ji} \equiv {a^6}\sum_{\mathbf{u},\mathbf {v}}\bar \zeta_j(\mathbf{u})\,\gamma_5 \zeta_i(\mathbf{v})  \, ,\label{eq:correlation-functs}
\end{align}
with the operators $A,P$ located in the bulk $(0<x_0<T)$ and the source operator ${{\mathcal{O}^{ji}}}$ located on the boundary $(x_0=0)$.
We also compute the correlation functions $g^{ij}_\mathrm{A,P}(T-x_0) $ with the same operator insertions in the bulk and sources
${\mathcal{O}^{\prime {ji} }}$ at $(x_0=T)$.
Due to the symmetric boundary conditions on the gauge fields, we can symmetrise $f^{ij}_\mathrm{A,P}$ and $g^{ij}_\mathrm{A,P}$, thus
reducing statistical fluctuations.
The renormalisation pattern and improvement constraint imply that the current (PCAC) mass $m_{ij}$, defined by 
\begin{align}   
&{m_{ij}(x_0)\equiv
\frac{\tilde\partial_0 f_\mathrm{A}^{ij}(x_0)+a c_\mathrm{A}\partial^\ast_0 \partial_0 f_\mathrm{P}^{ij}(x_0)}
{2\,f_\mathrm{P}^{ij}(x_0)}}  \, ,
\label{eq:pcac-mass}
\end{align}
can be parametrised as
\begin{align}   
m_{ij}(x_0) &= Z \, 
 \Big ( \cancel{{x} \, \mathrm{tr} \,\hat m^{({\rm sea})}} + \cancel{[{\color{red} z}+x \, ({\color{red}\bar{b}_\mathrm{A}}- {\color{red}\bar{b}_\mathrm{P}})]\, a\, (\mathrm{tr} \,\hat m^{({\rm sea})})^2}+ \cancel{{\color{red}y}\,a \,\mathrm{tr} \, \hat m^{2({\rm sea})}}
\label{eq:pcac-mass-parametrization}
\\
& + m_{q,ij}  \,(1+  [x \, \cancel{( {\color{red} b_\mathrm{A}} - {\color{red} b_\mathrm{P}}) + {\color{red} \bar b_\mathrm{m} } -( {\color{red} \bar b_\mathrm{A}} - {\color{red} \bar b_\mathrm{P}})] \, a\, \mathrm{tr} }\,\hat m^{({\rm sea})}  )
+ a m^2_{q,ij}  \, ( {\color{red} b_\mathrm{P}} - {\color{red} b_\mathrm{A}}) + \tfrac{1}{2} a  (m^2_{q,i}+m^2_{q,j}){\, \color{red}  b_\mathrm{m}} \Big ) \, ,\nonumber
\end{align}
where the slashed terms nearly vanish at ${\hat m^{({\rm sea})}\approx 0}$ and $Z$ indicates the ratio of renormalisation constants $Z(g_0^2)\equiv{Z_\mathrm{m}(g_0^2,{a/L})Z_\mathrm{P}(g_0^2, {a/L})}/{Z_\mathrm{A}(g_0^2)}$. For the various lattice derivatives standard notation is used: symmetric $\tilde\partial$, forward $\partial$, backward  $\partial^\ast$.{ Nearest-neighbour derivatives $\tilde\partial$ and  $\partial^\ast \partial$ suffer from $O(a^2)$ discretisation
errors; we label results produced with them with ``standard derivative''. In Refs.~\cite{deDivitiis:1997ka,Guagnelli:2000jw}, { next-to-nearest-neighbour} definitions have been proposed, with  $O(a^4)$ errors. Results obtained with these definitions are labelled with ``improved derivative''.
 
We determine the improvement coefficients  adopting the same strategy introduced 
for quenched QCD in~\cite{deDivitiis:1997ka,Guagnelli:2000jw,Heitger:2003ue,Bhattacharya:2000pn} and applied later for the two flavour case~\cite{Fritzsch:2010aw}. 
 We consider three different valence flavours $i,j=1,2,3$ and compute the four different PCAC masses 
 $m_{11}$,  $m_{22}$,  $m_{33}$,  $m_{12}$.  Up to renormalisation, these are physical quantities. We keep $m_{11}$ and $m_{22}$ fixed along our line of constant physics.
The hopping parameter $\kappa_1$ of the first valence flavour is set equal to the value of the dynamical quarks, in order to have nearly vanishig 
$m_{11}$. For the second valence flavour,
$\kappa_2$ is chosen so that $m_{22}$ is approximately equal to four arbitrary 
reference values:
 \begin{align}   
& L m_{11}\approx 0.0  \, ,\nonumber \\
&  L m_{22}\approx 0.25,\, 0.5,\, 0.75,\, 1.0 \,. \label{eq:constant-physics}
\end{align}
The third flavour is such that the corresponding bare mass is halfway the two others:
 \begin{align}   
&  m_{0,3}=\tfrac{1}{2}(m_{0,1}+m_{0,2})\,, \quad \mathrm{equivalently}  \quad m_{q,3}=\tfrac{1}{2}(m_{q,1}+m_{q,2}) \,.
\end{align}
The renormalisation and improvement structure of PCAC mass differences
is as follows:
\begin{align}
&\begin{cases}
 \Delta_{22,11}\equiv\tfrac{1}{2}\left( m_{22}- m_{11}\right)
&= Z \, \delta \; \left(  1+ {a A^{({\rm sea})}} +  
{\color{red}2 \, a\bar m }\, b_\mathrm{mAP} \right)+\dots \\ 
\Delta_{22,33}\equiv\left( m_{22}- m_{33}\right)
&= Z \, \delta \; \left(  1+ a A^{({\rm sea})} +  
{\color{red}(2 \, a\bar m + a\delta) }\, b_\mathrm{mAP}  \right) +\dots \\ 
\Delta_{33,11}\equiv\left( m_{33}- m_{11}\right)
&= Z \, \delta \; \left(  1+ a A^{({\rm sea})} +  
{\color{red}(2 \, a\bar m - a\delta) } \, b_\mathrm{mAP}  \right) +\dots \\ 
\Delta_{22,12}\equiv\left( m_{22}- m_{12}\right)
&= Z \, \delta \; \left(  1+ a A^{({\rm sea})} +  
{\color{red}2 \, a\bar m } \, b_\mathrm{mAP}   
{\color{red} -\, a\delta }\, b_\mathrm{AP} \right)+\dots \\ 
\Delta_{12,11}\equiv\left( m_{12}- m_{11}\right)
&= Z \, \delta \; \left(  1+ a A^{({\rm sea})} +  
{\color{red}2 \, a\bar m } \, b_\mathrm{mAP}   
{\color{red} +\, a\delta }\, b_\mathrm{AP} \right) +\dots\,, 
\end{cases} \label{eq:five}\\
& a\bar m \equiv \left(a m_{q,2}+a m_{q,1}\right)/2 \, , \qquad a\delta \equiv \left(a m_{q,2}-a m_{q,1}\right)/2  \, ,\nonumber\\
& 
a A^{({\rm sea})} \equiv (x \, { b_\mathrm{AP} +  \bar b_\mathrm{mAP}) \, a\, \mathrm{tr} }\,\hat m^{({\rm sea})}  \, ,\qquad b_\mathrm{mAP} \equiv  b_\mathrm{m}-( b_\mathrm{A}-b_\mathrm{P}), \qquad  b_\mathrm{AP} \equiv  b_\mathrm{A}-b_\mathrm{P}  \, .\nonumber
\end{align}
Both $a A^{({\rm sea})}$ and $Z$ cancel in the ratio of mass differences,  enabling us to single out  $b_\mathrm{A}-b_\mathrm{P}, b_\mathrm{m}$, as
well as $Z$:
 \begin{align}   
R_\mathrm{AP} 
&\equiv
\frac{\left(2 m_{12}- m_{11}- m_{22}\right)}
{\Delta \left(a m_{q,2}-a m_{q,1}\right)} =
{\color{red} b_\mathrm{A}-b_\mathrm{P}}+{O}(a m_{q,1}+a m_{q,2}) \, ,
\nonumber \\ 
R_\mathrm{m} 
&\equiv
\frac{2\left( m_{12}- m_{33}\right)}
{\Delta \left(a m_{q,2}-a m_{q,1}\right)} = 
\;{\color{red} b_\mathrm{m}} +{O}(a m_{q,1}+a m_{q,2}) \, ,
 \label{eq:main-results}\\ 
R_{Z} 
&\equiv 
\frac{ m_{11}- m_{22}}{ m_{q,1}- m_{q,2}}
+\left(R_\mathrm{AP}-R_\mathrm{m}\right)\left(a m_{11}+a m_{22}\right) = {\color{red} Z} + {O}(a\, \mathrm{tr} \,\hat m^{({\rm sea})}) \, .
\nonumber
\end{align}
In the above expressions, $\Delta$ without subscripts indicates any of the five $\Delta$'s in Eqs.~(\ref{eq:five}), leading to five possible determinations of the $b$'s, which differ by $O(a)$ terms. This ambiguity becomes $O(a^2)$ when the $b$'s are inserted in the definition of renormalised, improved
quark-masses. With exactly massless sea quarks the ambiguity in $Z$ is $O(a^2)$.
These formulae generalise the ones in previous works~\cite{deDivitiis:1997ka,Guagnelli:2000jw,Bhattacharya:2000pn,Fritzsch:2010aw}. 
%
\section{Simulation details}
\label{sec:details}
%
As already mentioned, our simulations are performed on a constant-physics trajectory in the space of bare parameters, with all physical scales held fixed, as illustrated in Fig.~\ref{fig:RGlattices}. 
We use the gauge configurations generated by the ALPHA collaboration, with the coupling constant $\beta=6/g_0^2$ tuned so that the physical lattice extent is fixed to $L\approx 1.2\,\mathrm{fm}$.
The tuning is based on the 2-loop perturbative expression for the lattice spacing. 
Subsequently, the value of $\kappa$ (corresponding to the mass of the degenerate sea quarks) is fixed for each lattice, so as to obtain a vanishing PCAC mass. 
The parameters of the available configurations are shown in Tab.~\ref{tab:parameters}. The values of $\beta$ span a range 
which is suitable for large-volume simulations. They correspond to the interval of lattice spacings $0.045 \,\mathrm{fm} \lesssim a \lesssim 0.090 \,\mathrm{fm}$. All lattices (except E1k1 and E1k2 where $T= 3L/2$) have temporal size $T= 3L/2 - a$. For details, see Ref.~\cite{Bulava:2015bxa,Bulava:2016ktf}.

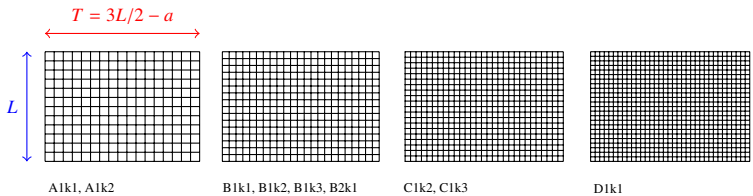
\begin{figure}[thb]
  \centering
\begin{tikzpicture}[scale=0.6]
\draw[step=.2cm]  (-1.8,-1.2001) grid (1.6,1.2); 
\draw[blue,<->] ({-2.2},{-1.2}) to node [left] {\scriptsize $L$} ({-2.2},{1.2});
\draw[red,<->] ({-1.8},{+1.6}) to node [above] {\scriptsize $T= 3L/2 - a$} ({1.6},{+1.6});
\draw[] (-1.,-1.8) node{\tiny A1k1, A1k2};
\draw[] (-1.+4.6,-1.8) node{\tiny B1k1, B1k2, B1k3, B2k1};
\draw[] (-1.+7.8,-1.8) node{\tiny C1k2, C1k3};
\draw[] (-1.+11.6,-1.8) node{\tiny D1k1};
\draw[step=.15cm] (-1.9001+4,-1.200) grid (1.55+4,1.2); 
\draw[step=.12cm] (-1.89+8.0,-1.200) grid (1.6+8.0,1.2); 
\draw[step=.1cm] (-1.8001+12,-1.201) grid (1.701+12,1.201); 
\end{tikzpicture}
\caption{Lattices with varying lattice spacing but identical physical size $L\approx 1.2\,\mathrm{fm}$.}
\label{fig:RGlattices}
\end{figure}
\begin{table}[thb]
  \small
 \sidecaption
\caption{
Simulation parameters $L,T,\beta,\kappa$, number of replicas  \#\,REP (i.e. number of statistically independent sets of configurations from Monte Carlo runs at identical parameters) and number of molecular
dynamics units \#\,MDU for each ensemble ID.}
  \label{tab:parameters}
    \centering
  \scalebox{0.85}{
\begin{tabular}{ccclccrrcc}\\
\hline 
$L^3\times T / a^4$ &&& $\beta$ && $\kappa$ & \#\,REP & \#\,MDU && ID   \\
\hline 
\hline 
 $12^3\times 17$     &&& 3.3       && 0.13652  & 10      & 10240   && A1k1 \\
		               &&&             && 0.13660  & 10      & 12620   && A1k2 \\
\hline 
 $14^3\times 21$    &&& 3.414  && 0.13690  & 32      & 10360   && E1k1 \\
		              &&&             && 0.13695  & 48      & 13984   && E1k2 \\\hline  
\color{red} $16^3\times 23$     &&& \color{red} 3.512   && 0.13700  & 2        & 20480   && B1k1 \\
                               &&&             && 0.13703  & 1       & 8192    && B1k2 \\
                               &&&             && \color{red} 0.13710  & \color{red} 3       & \color{red} 24560   && \color{red} B1k3 \\
\hline 
$16^3\times 23$     &&& 3.47    && 0.13700  & 3       & 29584    && B2k1 \\
\hline
$20^3\times 29$     &&& 3.676   && 0.13700  & 4       & 15232   && C1k2 \\ 
                               &&&             && 0.13719  & 4       & 15472   && C1k3 \\
\hline 
$24^3\times 35$     &&& 3.810   && 0.13712  & 5       & 10240   && D1k1 \\
\hline 
\end{tabular}
}
\end{table}
The SF simulations have been performed using the \texttt{openQCD} code~\cite{openQCD},  with improved L\"uscher--Weisz gauge action~\cite{Luscher:1984xn}, $N_f=3$ massless Wilson-clover fermions, vanishing boundary gauge fields $C=C'=0$  and boundary fermion parameter
$\theta=0$. The value of the improvement coefficient $c_\mathrm{SW}$ is taken from Ref.~\cite{Bulava:2013cta}.
The RHMC algorithm~\cite{Kennedy:1998cu,Clark:2006fx,Luscher:2008tw} is used for the third dynamical quark.
%
\section{Results}
\label{sec:results}
%
The preliminary results presented in the this work are obtained from the analysis of the B1k3 ensemble,
marked in red in Tab.~\ref{tab:parameters}.

The  time dependence of the PCAC masses $m_{11}$,  $m_{22}$,  $m_{33}$,  $m_{12}$ 
is shown in Fig.~\ref{fig:massesB1k3}. These results are obtained with improved derivatives; those obtained
with standard derivatives do not show appreciable differences.  All masses show wide plateaux, and the statistical errors are smaller
than the symbols.
The red points correspond to the chiral flavour $m_{11}\approx 0$, while the blue data represent $m_{22}$. As can be seen on the right vertical axis, $m_{22}$ is tuned with good precision to the chosen reference values $L m_{22} \approx 0.25, \, 0.5, \, 0.75,\, 1.0$ of Eq.~(\ref{eq:constant-physics}).
\begin{figure}[thb]
  \centering
  \includegraphics[width=9cm,clip]{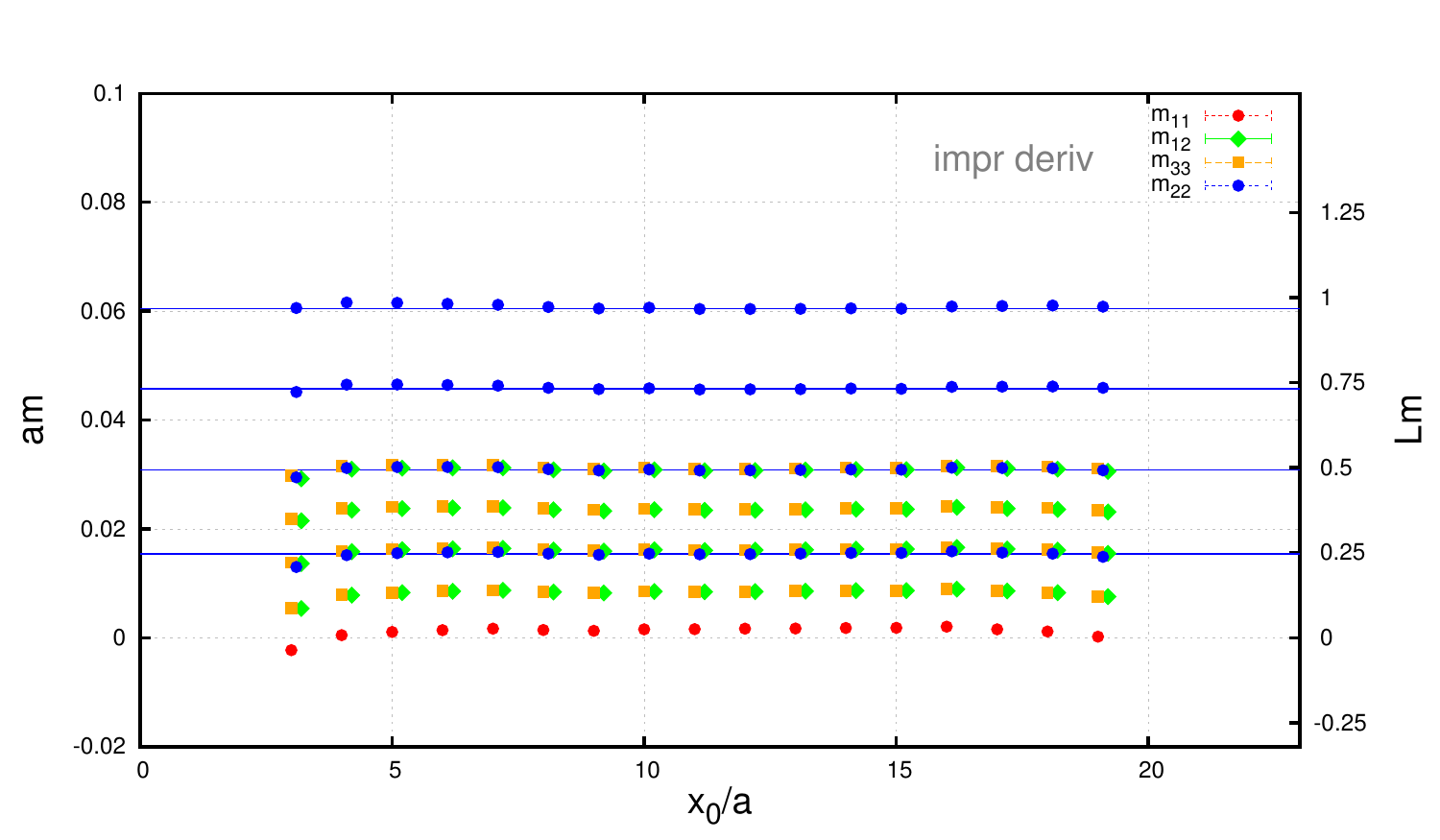}
  \caption{Time dependence of PCAC masses for the ensemble B1k3.}
  \label{fig:massesB1k3}
\end{figure}

To check the consistency of our data with the parametrisation of the cutoff effects given in Eqs.~(\ref{eq:five}),
we verify  that the quantities
\begin{align}
r_1&\equiv \frac{1}{4}\,\frac{\left( m_{22}- m_{11}\right) \left( m_{22}- m_{11}\right)}{\left( m_{22}- m_{33}\right) \left( m_{33}- m_{11}\right)}-1= {\color{red}O(a^2)}  \, ,\nonumber \\
r_2&\equiv\frac{1}{4}\,\frac{\left( m_{22}- m_{11}\right) \left( m_{22}- m_{11}\right)}{\left( m_{22}- m_{12}\right) \left( m_{12}- m_{11}\right)}-1= {\color{red}O(a^2)} \, ,
\label{eq:ratios}
\end{align}
are close to zero. As can be seen in Fig.~\ref{fig:ratios}, these ratios are of order $10^{-4}$ and less, significantly smaller than   
the values $a m_{q,2}\approx 0.015, \,0.06 $, with improved-derivative data having the smaller values. Moreover, they tend to increase with the mass $m_{22}$ and time $x_0$, as expected.
The smallness of  $r_1$ and $r_2$ demonstrates that results for the $b$'s are insensitive to the choice of  $\Delta$  in the denominator.
In what follows we set $\Delta = \Delta_{22,11}$, which is the one kept fixed on the line of constant physics.
\begin{figure}[t]
   \centering
   \subfigure[$r_1.$]%
             {\includegraphics[width=\textwidth,clip]{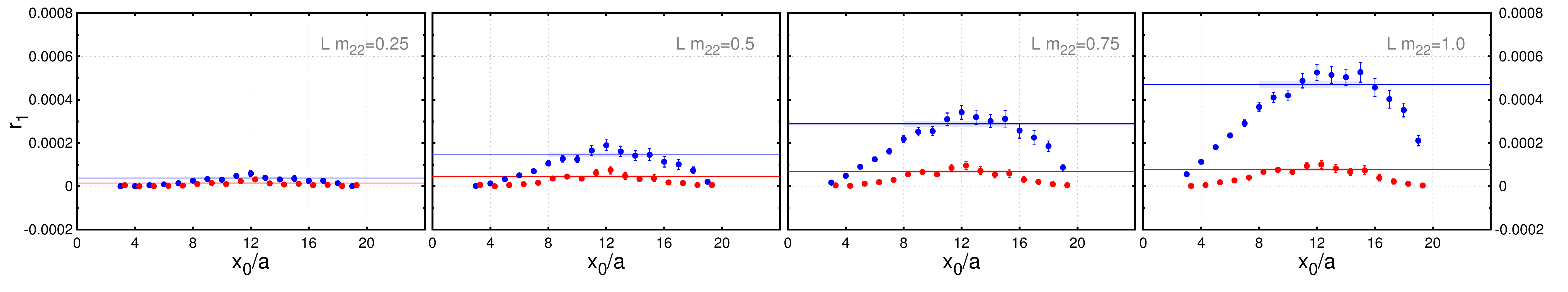}}\hfill
   \subfigure[$r_2.$]%
             {\includegraphics[width=\textwidth,clip]{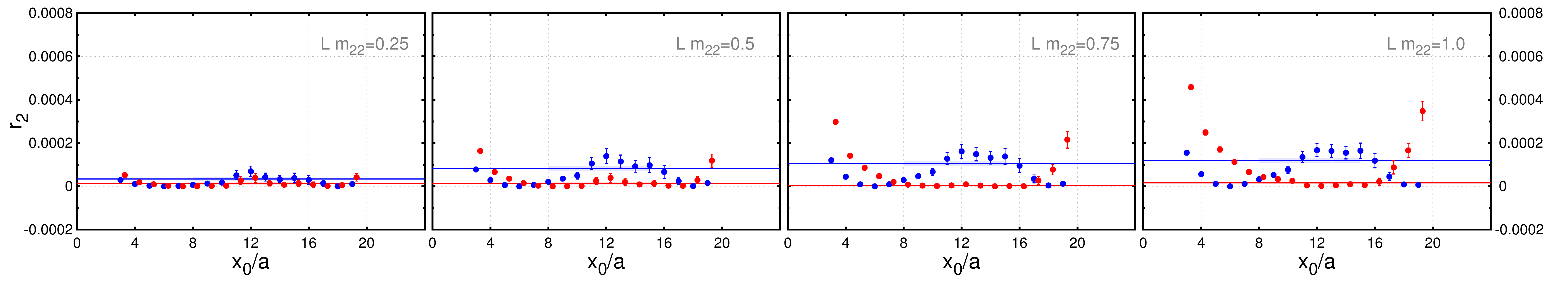}}\hfill
   \caption{Time dependence of the ratios $r_1$ and $r_2$ for the ensemble B1k3.    Blue points refer to PCAC masses computed with standard derivatives, red points to those computed with improved ones.
   The four plots correspond to the four reference values $Lm_{22}=0.25, \, 0.5,\, 0.75,\, 1.0$.}
   \label{fig:ratios}
\end{figure}

The main results of our preliminary analysis are presented in Fig.~\ref{fig:mainresults}. The plots (a),(b) and (c)
show the time dependence of  estimators for $b_\mathrm{AP}$, $b_\mathrm{m}$ and  $Z$, respectively, with
 blue points corresponding to the standard derivative and red points to the improved one.
The horizontal lines in the plots indicate the averages over the time window $x_0/a = [8; 15]$, corresponding to the middle third of the time extent $T$. Averaging over time slices is part of our operative definition of the parameters $b_\mathrm{AP}$, $b_\mathrm{m}$, $Z$.
 Note that
$R_\mathrm{AP}$ data show a significant ambiguity with respect to the choice of the lattice derivative, as previously observed in the  quenched and $N_f=2$ studies~\cite{deDivitiis:1997ka,Guagnelli:2000jw,Fritzsch:2010aw}.

In general all signals show better plateaux and smaller statistical errors at larger values of $m_{22}$, where however discretisation effects
are expected to be larger.

\begin{figure}[tp]
   \centering
   \subfigure[$R_\mathrm{AP}$.]%
            {\includegraphics[width=\textwidth,clip]{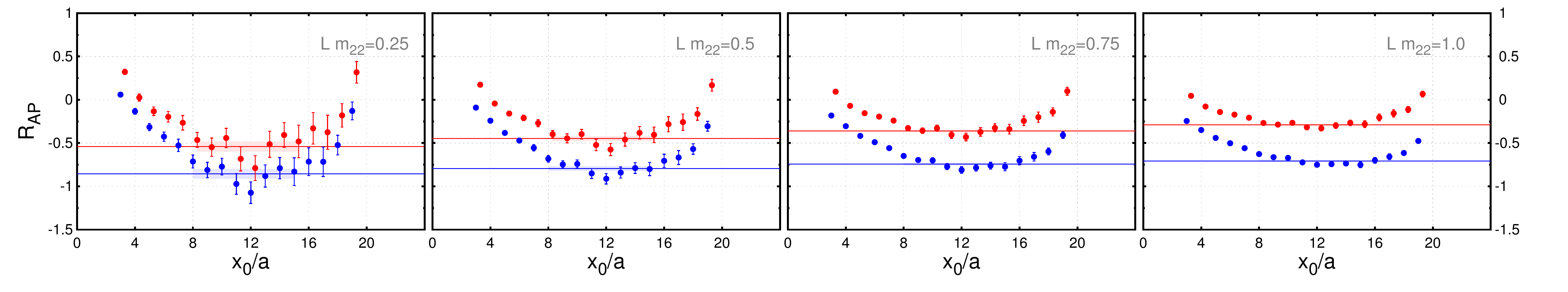}}\hfill
   \subfigure[$R_\mathrm{m}$.]%
             {\includegraphics[width=\textwidth,clip]{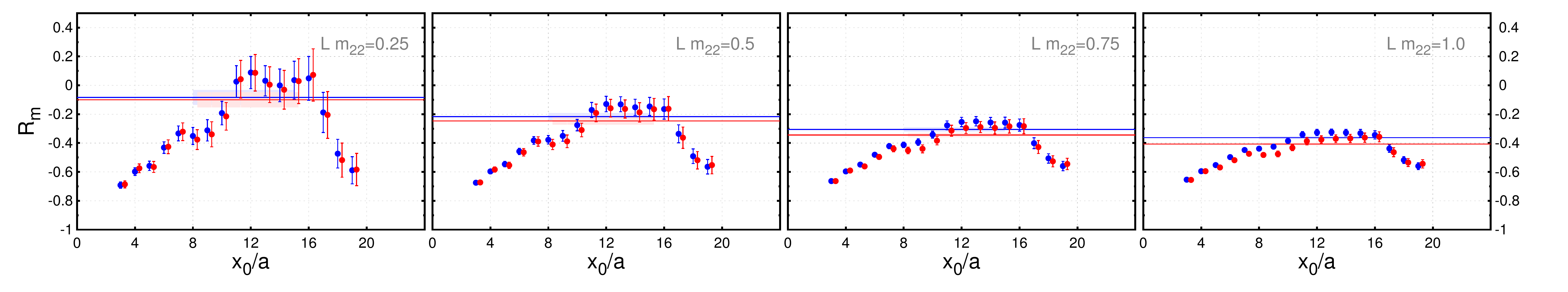}}
      \subfigure[$R_Z$.]%
             {\includegraphics[width=\textwidth,clip]{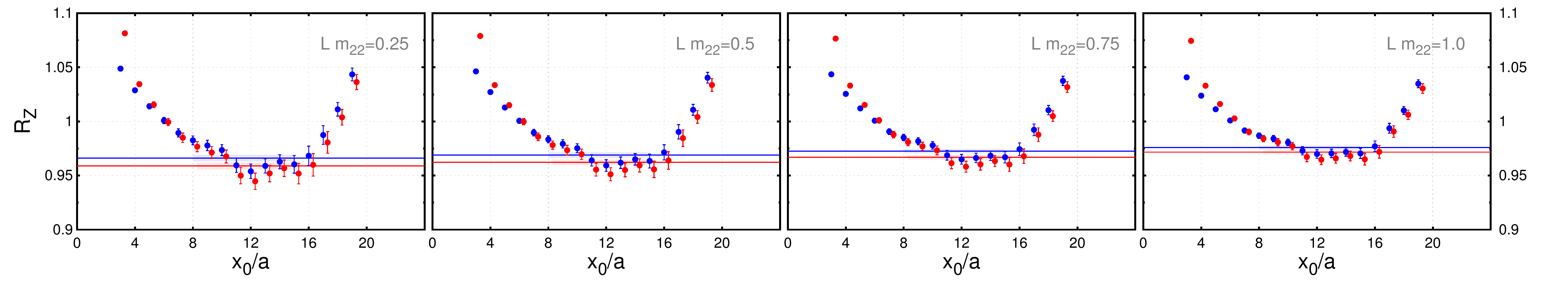}}\hfill           
    \caption{Time dependence of $R_\mathrm{AP}$, $R_\mathrm{m}$ and $R_Z$ for the ensemble B1k3. Blue points refer to PCAC masses computed with standard derivatives, red points to those computed with improved ones. The four plots correspond to the four reference values $Lm_{22}=0.25, \, 0.5,\, 0.75,\, 1.0$. }
   \label{fig:mainresults}
\end{figure}
%
\subsection{Topological sectors}
\label{sec:topological}
Since Ward identities hold in any topological sector and the improvement coefficients are 
short distance quantities, our results should be insensitive to the topological charge $Q$. 
Following Ref.~\cite{Bulava:2015bxa}, we repeated our data analysis only considering configurations belonging to the trivial (i.e. $Q=0$) topological sector, using a topological charge defined through 
gradient-flow fields~\cite{Luscher:2010iy,Luscher:2011bx}
 \begin{align}
Q(t)&\equiv-\frac{a^4}{32\pi^2}\sum_x\, \epsilon_{\mu\nu\alpha\beta}\, \mathrm{tr} \{  G_{\mu\nu}(x,t) \,  G_{\alpha\beta}(x,t) \}\,,\qquad\qquad G_{\mu \nu} \equiv \partial_\mu B_\nu -\partial_\nu B_\mu + [B_\mu ,B_\nu]\,,
\label{eq:topological-charge}
\end{align}
where $t$ is the flow time, kept fixed in units of physical volume, and $B_\mu$ is the gluon field.
The results were in agreement with the full statistics (i.e. including all topological charges), while only reflecting fluctuations consistent with the reduction of statistics. This confirms the aforementioned expectation of the results' insensitivity to topology.
%
\section{Conclusion}
\label{sec:conclusion}  
%
To complete our work we will compute the correlation functions for full statistics and on all available lattices
at different lattice spacings (see Tab.~\ref{tab:parameters}). Combining the known analytic perturbative expressions
for these quantities, valid towards vanishing $g_0^2$, with our data points, we aim at obtaining suitable interpolation functions
for $b_\mathrm{AP}(g_0^2)$, $b_\mathrm{m}(g_0^2)$, and $Z(g_0^2)$. These non-perturbative formulae are needed  for
reaching $O(a)$ improved results in simulations of lattice QCD with $N_f = 3$ Wilson quarks in large volume.
It will be interesting to compare our results to those recently obtained by Korcyl and Bali~\cite{Korcyl:2016ugy}, using a different non-perturbative renormalisation method.
%
\section*{Acknowledgements}
\label{sec:acknowledgements}  
%
Computer resources were provided by the INFN (GALILEO cluster at CINECA) and the ZIV of the University of M\"{u}nster (PALMA HPC cluster).
This work was supported by the grant {HE~4517/3-1} (J.~H.) of the Deutsche Forschungsgemeinschaft.
C.~C.~K., scholar of the German Academic Scholarship Foundation (Studienstiftung des deutschen Volkes), gratefully acknowledges their financial and academic support.
 \emph{
 The speaker wishes to thank Isabel Campos, Elvira G\'amiz and all the organizers 
of the 2017 lattice conference for their wonderful hospitality in Granada.}
\bibliography{lattice2017}


\end{document}